# Quasi-Cyclic Asymptotically Regular LDPC Codes

David G. M. Mitchell*, Roxana Smarandache†, Michael Lentmaier‡, and Daniel J. Costello, Jr.*

*Dept. of Electrical Engineering, University of Notre Dame, Notre Dame, Indiana, USA,
{david.mitchell, costello.2}@nd.edu

†Dept. of Mathematics and Statistics, San Diego State University, San Diego, California, USA,
rsmarand@sciences.sdsu.edu

‡Vodafone Chair Mobile Communications Systems, Dresden University of Technology, Dresden, Germany,
michael.lentmaier@ifn.et.tu-dresden.de

***Abstract*— Families of *asymptotically regular* LDPC block code ensembles can be formed by terminating $(J,K)$-regular protograph-based LDPC convolutional codes. By varying the termination length, we obtain a large selection of LDPC block code ensembles with varying code rates, minimum distance that grows linearly with block length, and capacity approaching iterative decoding thresholds, despite the fact that the terminated ensembles are almost regular. In this paper, we investigate the properties of the quasi-cyclic (QC) members of such an ensemble. We show that an upper bound on the minimum Hamming distance of members of the QC sub-ensemble can be improved by careful choice of the component protographs used in the code construction. Further, we show that the upper bound on the minimum distance can be improved by using arrays of circulants in a graph cover of the protograph.**

## I. Introduction

Low-density parity-check (LDPC) codes [1] based on a *protograph* [2] form a subclass of multi-edge type codes that have been shown to have many desirable features, such as good iterative decoding thresholds and, for suitably-designed protographs, linear minimum distance growth (see, e.g., [3]). Analogously, ensembles of LDPC convolutional codes [4], the convolutional counterparts to LDPC block codes, can also be constructed using protographs and display the same desirable properties (see [5] and [6], respectively).

So-called *asymptotically regular* LDPC block code ensembles [7] are formed by terminating $(J,K)$-regular protograph-based LDPC convolutional codes. This construction method results in LDPC block code ensembles with substantially better thresholds than those of $(J,K)$-regular LDPC block code ensembles, despite the fact that the ensembles are almost regular (see, e.g., [7]). These codes were analysed further in [8] and were also shown to have minimum distance growing linearly with block length, i.e., they are asymptotically good. As the termination length tends to infinity, it is further observed that the iterative decoding thresholds of these asymptotically good ensembles approach the optimal maximum a posteriori probability (MAP) decoding thresholds of the corresponding LDPC block code ensembles. More recently, this property has been proven analytically in [9] for the binary erasure channel (BEC) considering some slightly modified ensembles.

Members of the protograph-based LDPC code ensemble that are quasi-cyclic (QC) are of great interest to code designers, since they can be encoded with low complexity using simple feedback shift-registers [10]. Moreover, QC codes can be shown to perform well compared to random codes for moderate block lengths [11], [12]. However, unlike typical members of an asymptotically good protograph-based LDPC code ensemble, codes from the QC sub-ensemble cannot be asymptotically good. Indeed, if the protograph base matrix consists of only ones and zeros, then the minimum Hamming distance is immediately bounded above by $(n_c+1)!$, where $n_c$ is the number of check nodes in the protograph [13], [14].

In this paper, building on recent results by Smarandache and Vontobel [15], we show that the upper bound on the minimum Hamming distance of members of the QC sub-ensemble of asymptotically regular $(J,K)$ LDPC codes can be improved by careful choice of the component protographs used in the code construction. Even though we show that the QC codes from the ensemble are not 'typical', we see that constructions that improve the ensemble minimum distance growth rate also increase the upper bound on minimum distance for members of the QC sub-ensemble. In addition, for several of the examples given in the paper, QC codes are constructed that achieve this upper bound. Further, we show that the upper bound on minimum distance can be improved by using arrays of circulants in a graph cover of the protograph.

This work was partially supported by NSF Grants CCF08-30650, DMS-0708033, and TF-0830608 and NASA Grant NNX09AI66G.

## II. Analysis of protograph-based LDPC codes

A protograph is a small bipartite graph $B=(V,C,E)$ that connects a set of $n_v$ variable nodes $V=\{v_0,\ldots,v_{n_v-1}\}$ to a set of $n_c$ check nodes $C=\{c_0,\ldots,c_{n_c-1}\}$ using a set of edges $E$. The protograph can be represented by a parity-check or *base* biadjacency matrix $\mathbf{B}$, where $B_{x,y}$ is taken to be the number of edges connecting variable node $v_y$ to check node $c_x$. Figure 1 shows an example of an irregular protograph with repeated edges and the associated base matrix.

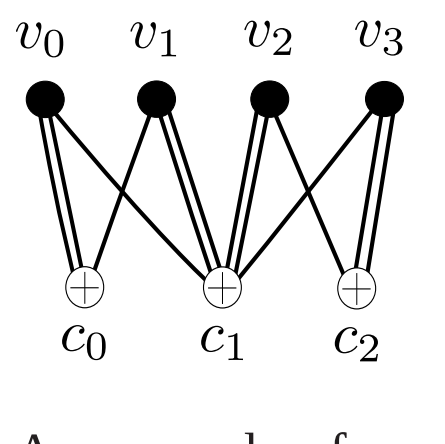

$$\mathbf{B} = \begin{bmatrix} 2 & 1 & 0 & 0 \\ 1 & 2 & 2 & 1 \\ 0 & 0 & 1 & 2 \end{bmatrix}$$

Fig. 1: An example of a protograph and the associated base matrix.

This protograph is called irregular because both the variable and check node degrees are not constant.

An ensemble of protograph-based LDPC block codes can be created from a base matrix $\mathbf{B}$ using a *copy-and-permute*

operation [2]. A parity-check matrix $\mathbf{H}$ from the ensemble of protograph-based LDPC block codes can then be obtained by replacing ones with an $N \times N$ permutation matrix and zeros with the $N \times N$ all zero matrix in the base matrix $\mathbf{B}$. In the case when a variable node and a check node are connected by $r$ repeated edges, the associated entry in $\mathbf{B}$ equals $r$ and the corresponding block in $\mathbf{H}$ consists of a summation of $r$ $N \times N$ permutation matrices. The *ensemble* is defined as the set of all possible parity-check matrices $\mathbf{H}$ that can be formed using this method.

By construction, every code in the resulting ensemble has the same node degrees and structure. The ensemble design rate is given as $R = 1 - n_c/n_v$. In addition, the sparsity condition of an LDPC matrix is satisfied for large $N$. The code created by applying the copy-and-permute operation to an $n_c \times n_v$ protograph base matrix $\mathbf{B}$ has block length $n = Nn_v$.

### A. Density evolution for protograph-based ensembles

Since every member of the protograph-based ensemble preserves the structure of the base protograph, density evolution analysis for the resulting codes can be performed within the protograph. In this paper, we assume that belief propagation (BP) decoding is performed after transmission over a BEC with erasure probability $\varepsilon$. Let $p^{(i)}$ denote the probability that the incoming message in the previous update along an edge of an arbitrary check node is an erasure. Then the *density evolution threshold* of an ensemble is defined as the maximal value of the channel parameter $\varepsilon$ for which $p^{(i)}$ converges to zero for all edges as the number of iterations $i$ tends to infinity.

### B. Weight enumeration for protograph-based ensembles

The preserved structure of members of a protograph-based LDPC code ensemble also facilitates the calculation of average weight enumerators. An *ensemble average weight enumerator* $A_d$ tells us that, given a particular Hamming weight $d$, a typical member of the ensemble has $A_d$ codewords with Hamming weight $d$. Combinatorial techniques for calculating enumerators for protograph-based ensembles have been presented in [3] and [16]. The weight enumerator $A_d$ can be analysed asymptotically to test if the ensemble is asymptotically good. If this is the case, then we can say that the majority of codes in the ensemble have minimum distance growing linearly at least as fast as $n\delta_{min}$, where $\delta_{min}$ is the *minimum distance growth rate* of the code ensemble [3].

## III. Quasi-cyclic protograph-based LDPC codes

One of the main advantages of quasi-cyclic LDPC codes is that they can be described simply, and as such are attractive for implementation purposes (see, e.g., [10]). In this section, we focus on the quasi-cyclic sub-ensembles of protograph-based ensembles of LDPC codes and review the existing literature that will be used to analyse these ensembles.

### A. Structure of QC sub-ensembles

Given a protograph base matrix $\mathbf{B}$, a parity-check matrix $\mathbf{H}$ from the ensemble of protograph based codes $\xi_{\mathbf{B}}(N)$ is created by replacing each non-zero entry $r$ with a summation of $r$ non-overlapping permutation matrices of size $N \times N$ and replacing zeros with the $N \times N$ all-zero matrix. The quasi-cyclic sub-ensemble, denoted $\xi_{\mathbf{B}}^{QC}(N)$, is the subset of parity-check matrices from $\xi_{\mathbf{B}}(N)$, where each of the permutation submatrices are chosen to be *circulant*. The notation $I_a$ is used to denote the $N \times N$ identity matrix with each row cyclically shifted to the left by $a$ positions. The set of all such matrices comprise the circulant subset of the set of $N \times N$ permutation matrices. When applying the copy-and-permute operation, by restricting the choice of permutation matrices to come from this subset, the resulting parity-check matrix $\mathbf{H}$ will be quasi-cyclic, i.e., $\mathbf{H} \in \xi_{\mathbf{B}}^{QC}(N) \subseteq \xi_{\mathbf{B}}(N)$. For example, a quasi-cyclic parity-check matrix can be formed from the base matrix defined in Figure 1 as

$$\mathbf{H} = \begin{bmatrix} I_1 + I_2 & I_4 & 0 & 0 \\ I_5 & I_{10} + I_{20} & I_9 + I_{18} & I_7 \\ 0 & 0 & I_{11} & I_{23} + I_{17} \end{bmatrix} \in \xi_{\mathbf{B}}^{QC}(N).$$

When considering a sub-ensemble such as $\xi_{\mathbf{B}}^{QC}(N)$, one has to be careful with the relevance of asymptotic results obtained for the ensemble $\xi_{\mathbf{B}}(N)$. As $N \to \infty$, if the probablility of choosing a member of the sub-ensemble is non-zero we say that the code is a *typical* member of the ensemble. By this definition, it is clear that the sub-ensemble $\xi_{\mathbf{B}}^{QC}(N)$ contains atypical codes. This follows since there are only $N$ out of $N!$ permutations that are circulant, i.e., the fraction of choices of permutation matrices that are circulant is $N/N! = 1/(N-1)!$, which tends to zero as $N \to \infty$. Then, if the base matrix $\mathbf{B}$ contains only ones and zeros, the fraction of codes in the ensemble that are circulant is $(1/(N-1)!)^k$, where $k$ is the number of ones in $\mathbf{B}$. Repeated edges in $\mathbf{B}$ further reduce this fraction.

### B. Minimum Hamming distance bounds for QC sub-ensembles

If the base matrix $\mathbf{B}$ contains only ones and zeros, then it is well known that the minimum Hamming distance of any code from the quasi-cyclic sub-ensemble of protograph-based LDPC codes can immediately be bounded above by $(n_c+1)!$ [13], [14]. This result was improved and extended by Smarandache and Vontobel to base matrices with entries larger than one [15]. Let the *permanent* of an $m \times m$ matrix $\mathbf{B}$ be defined as

$$\text{perm}(\mathbf{B}) = \sum_{\sigma} \prod_{x=1}^{m} B_{x,\sigma(x)},$$

where we sum over the $m!$ permutations $\sigma$ of the set $\{1, \ldots, m\}$. Then the minimum distance of a code drawn from the QC sub-ensemble can be upper bounded as follows:

*Theorem 1:* Let $C$ be a code from $\xi_{\mathbf{B}}^{QC}(N)$, the quasi-cyclic sub-ensemble of the protograph-based ensemble of codes formed from base matrix $\mathbf{B}$. Then the minimum Hamming distance of $C$ is bounded above as[1]

$$d_{min}(C) \leq \min^*_{\substack{S \subseteq \{1,\ldots,n_v\} \\ |S| = n_c+1}} \sum_{i \in S} \text{perm}(\mathbf{B}_{S \setminus i}), \quad (1)$$

where $\text{perm}(\mathbf{B}_{S\setminus i})$ denotes the permanent of the matrix constructed as the $n_c$ columns of $\mathbf{B}$ from the set $S \setminus i$.

[1]The $\min^*\{\cdot\}$ operator returns the smallest non-zero value from a set. In this context, if the all-zero codeword arises from a constructed matrix, this operator ensures that 0 is disregarded as an upper bound in the minimization.

## IV. Terminated protograph-based LDPC convolutional codes

A rate $R = b/c$ (time-varying) binary LDPC convolutional code [4] can be defined as the set of infinite binary sequences $\mathbf{v}_{[-\infty,\infty]}$ that satisfy the equation $\mathbf{v}_{[-\infty,\infty]}\mathbf{H}^{\mathrm{T}}_{[-\infty,\infty]} = \mathbf{0}$, where

$$\mathbf{H}^T_{[-\infty,\infty]} = \begin{bmatrix} \ddots & & \ddots & & \\ \mathbf{H}_0^T(0) & \cdots & \mathbf{H}_{m_s}^T(m_s) & & \\ & \ddots & & \ddots & \\ & & \mathbf{H}_0^T(t) & \cdots & \mathbf{H}_{m_s}^T(t+m_s) \\ & & \ddots & & \ddots \end{bmatrix}$$

is the transposed parity-check matrix, also called the *syndrome former matrix*. The binary $(c-b)\times c$ submatrices $\mathbf{H}_i(t)$, $i = 0, 1, \cdots, m_s$, satisfy the conditions that $\mathbf{H}_{m_s}(t) \neq \mathbf{0}$ for at least one $t \in \mathbb{Z}$ and that $\mathbf{H}_0(t)$ has full rank for all $t$. We call $m_s$ the *syndrome former memory* and $\nu_s = (m_s+1)\cdot c$ the *decoding constraint length*. These parameters determine the width of the nonzero diagonal region of $\mathbf{H}_{[-\infty,\infty]}$. The sparsity of the parity-check matrix is insured by demanding that its rows have Hamming weight much less than $\nu_s$. The code is said to be regular if its parity-check matrix $\mathbf{H}_{[-\infty,\infty]}$ has exactly $J$ ones in every column and $K$ ones in every row.

### A. Constructing protograph-based LDPC convolutional codes

Analogously to block codes, an ensemble of LDPC convolutional codes can be constructed from a protograph. We proceed by forming a time-invariant infinite base matrix[2] with component $b_c \times b_v$ submatrices $\mathbf{B}_0, \mathbf{B}_1, \ldots, \mathbf{B}_{m_s}$ as follows:

$$\mathbf{B}_{[-\infty,\infty]} = \begin{bmatrix} \ddots & & \ddots & & \\ \mathbf{B}_{m_s} & \cdots & \mathbf{B}_0 & & \\ & \ddots & & \ddots & \\ & & \mathbf{B}_{m_s} & \cdots & \mathbf{B}_0 \\ & & \ddots & & \ddots \end{bmatrix}. \quad (2)$$

The infinite Tanner graph associated with $\mathbf{B}_{[-\infty,\infty]}$ can be regarded as a *convolutional protograph*. An ensemble of time-varying LDPC convolutional codes can be formed from $\mathbf{B}_{[-\infty,\infty]}$ using the protograph construction method based on $N \times N$ permutation matrices described in Section II. Given a base matrix $\mathbf{B}$, one can form a convolutional protograph with the same rate and degree distribution by creating the submatrices $\mathbf{B}_0, \mathbf{B}_1, \ldots, \mathbf{B}_{m_s}$ using an *edge-spreading* technique [7]. Here, the edges of the protograph base matrix $\mathbf{B}$ are spread over the component submatrices such that $\mathbf{B}_0 + \mathbf{B}_1 + \ldots + \mathbf{B}_{m_s} = \mathbf{B}$. Note that the submatrices necessarily have the same size as $\mathbf{B}$.

### B. Forming terminated protograph-based LDPC convolutional codes

Suppose that we start the base matrix defined in (2) at time $t = 0$ and terminate it after $L$ time instants. The resulting finite-length base matrix is given by

[2]If the base matrix contains only ones and zeros, it represents the parity-check matrix of a rate $R = 1 - b_c/b_v$ time-invariant convolutional code with syndrome former memory $m_s$.

$$\mathbf{B}_{[0,L-1]} = \begin{bmatrix} \mathbf{B}_0 & & \\ \vdots & \ddots & \\ \mathbf{B}_{m_s} & & \mathbf{B}_0 \\ & \ddots & \vdots \\ & & \mathbf{B}_{m_s} \end{bmatrix}_{(L+m_s)b_c \times Lb_v}. \quad (3)$$

The matrix $\mathbf{B}_{[0,L-1]}$ can be considered as the base matrix of a terminated protograph-based LDPC convolutional code ensemble. Termination in this fashion results in a rate loss. The design rate $R_L$ of the terminated code ensemble is equal to

$$R_L = 1 - \left(\frac{L+m_s}{L}\right)\frac{b_c}{b_v} = 1 - \left(\frac{L+m_s}{L}\right)(1-R), \quad (4)$$

where $R = 1 - Nb_c/Nb_v = 1 - b_c/b_v$ is the rate of the unterminated LDPC convolutional code ensemble. Note that, as the termination factor $L$ increases, the rate increases and approaches the rate of the unterminated LDPC convolutional code ensemble. In addition, as $L \to \infty$, the degree distribution approaches that of the unterminated ensemble. It follows that if the base matrix $\mathbf{B}$ is $(J,K)$-regular, and we apply the edge spreading technique to preserve the structure, the degree distribution of the terminated ensemble approaches that of a $(J,K)$-regular ensemble as $L \to \infty$, i.e., it is asymptotically regular. The protograph-based LDPC block code ensemble associated with $\mathbf{B}_{[0,L-1]}$ can be studied using the analysis discussed in Section II.

## V. QC asymptotically regular LDPC codes

In this section, we form families of asymptotically regular LDPC block code ensembles by terminating $(J,K)$-regular protograph-based LDPC convolutional codes. It was shown in [8] that the minimum distance growth rates and the iterative decoding thresholds of asymptotically good terminated ensembles are sensitive to the choice of component protographs used in the edge spreading technique. Here, we investigate how the choice of component protographs affects the upper bound on the minimum Hamming distance of the QC sub-ensemble $\xi^{QC}_{\mathbf{B}_{[0,L-1]}}(N)$. Even though the QC codes are not typical members of the ensemble, we observe that choosing component submatrices that yield strong ensemble minimum distance growth rates also gives large upper bounds on the minimum distance of the QC sub-ensemble. Further, we show that by using arrays of circulants, which can alternatively be viewed as the QC sub-ensemble arising from a graph-cover of the protograph, we can increase the upper bound on the Hamming distance of codes chosen from $\xi^{QC}_{\mathbf{B}_{[0,L-1]}}(N)$.

To begin, we compare different edge spreadings that result in asymptotically regular $(3,6)$ ensembles.

*Example* 1: Consider spreading the edges of the base matrix $\mathbf{B} = \begin{bmatrix} 3 & 3 \end{bmatrix}$ into component submatrices

$$\mathbf{B}_0 = \begin{bmatrix} 1 & 1 \end{bmatrix} = \mathbf{B}_1 = \mathbf{B}_2,$$

where $\mathbf{B}_0 + \mathbf{B}_1 + \mathbf{B}_2 = \mathbf{B}$. Using these component submatrices, we can obtain the base matrix for a $(3,6)$-regular LDPC convolutional code ensemble with syndrome former memory $m_s = 2$. The terminated ensembles in this family were shown to be asymptotically good with thresholds converging to the (optimal) MAP decoding threshold $\varepsilon^* = 0.4881$ for $(3,6)$-regular LDPC codes on the BEC as $L \to \infty$ [8]. For termination factor $L = 4$, the ensemble has design rate $R_4 = 1/4$, minimum distance growth rate $\delta_{min}^{(4)} = 0.0814$, and BEC iterative decoding threshold $\varepsilon^* = 0.635$. Terminating after $L = 10$ time instants, the rate increases to $R_{10} = 2/5$, the minimum distance growth rate is $\delta_{min}^{(10)} = 0.0258$, and the threshold is $\varepsilon^* = 0.505$. As $L \to \infty$, the minimum distance growth rate tends to zero and the threshold converges to $\varepsilon^* = 0.488$ (close to the Shannon limit $\varepsilon_{sh} = 0.5$ for rate $R_\infty = 1/2$).

Using Theorem 1 and the base matrix $\mathbf{B}_{[0,2]}$ $(L = 3)$, we calculate that the minimum Hamming distance for the circulant sub-ensemble $\xi^{QC}_{\mathbf{B}_{[0,2]}}(N)$ is bounded above by 56, i.e., $d_{minQC} \leq 56$ for any circulant size $N$. To show that this upper bound is indeed achievable, consider the following parity-check matrix:

$$\mathbf{H} = \begin{bmatrix} I_1 & I_2 & 0 & 0 & 0 & 0 \\ I_5 & I_{10} & I_{20} & I_9 & 0 & 0 \\ I_{25} & I_{19} & I_7 & I_{14} & I_{28} & I_{11} \\ 0 & 0 & I_4 & I_8 & I_{16} & I_{22} \\ 0 & 0 & 0 & 0 & I_{18} & I_{34} \end{bmatrix} \in \xi^{QC}_{\mathbf{B}_{[0,2]}}(N).$$

With circulant size $N = 49$, this parity-check matrix defines a $[294, 51, 56]$ QC binary linear code with girth 8 (in this case, $\mathbf{H}$ has 2 redundant rows). Note that, for typical codes from the ensemble $\xi_{\mathbf{B}_{[0,2]}}(N)$, the (asymptotic) minimum distance growth rate is $\delta_{min}^{(3)} = 0.1419$.

For termination factors $L > 3$, the upper bound $d_{minQC} \leq 56$ remains constant. It follows that this is also an upper bound on the free distance of the circulant sub-ensemble of protograph-based LDPC convolutional codes, i.e., $d_{freeQC} \leq 56$. In addition, as the termination length of the convolutional protograph increases, the asymptotically regular ensembles display capacity approaching iterative decoding thresholds. Even though these thresholds are not achievable with QC codes because small cycles exist in the Tanner graph, we expect that QC codes drawn from ensembles with a better iterative decoding threshold will display better performance in the waterfall region of the bit error rate curve, even for finite block lengths (see, e.g., [17]). In practice, the design parameter $L$ adds an additional degree of freedom to existing block code designs. Starting from any LDPC block code, it is possible to derive terminated convolutional codes that share the same encoding and decoding architecture for arbitrary $L$.

*Example* 2: Let $\mathbf{B}$ be the all-ones matrix of size $3 \times 6$. Consider the following edge spreading of $\mathbf{B}$:

$$\mathbf{B}_0 = \begin{bmatrix} 1 & 1 & 1 & 0 & 0 & 0 \\ 0 & 1 & 1 & 1 & 0 & 0 \\ 0 & 0 & 0 & 1 & 1 & 1 \end{bmatrix} \text{ and } \mathbf{B}_1 = \mathbf{B} - \mathbf{B}_0.$$

Using $\mathbf{B}_0$ and $\mathbf{B}_1$ as given above, the asymptotically regular $(3,6)$ ensemble defined by (3) has six degree 3 check nodes and $3L - 3$ degree 6 check nodes for termination factors $L \geq 2$. The protographs in this terminated family will be highly regular with no degree 2 check nodes. The family of terminated $(3,6)$-regular LDPC convolutional code ensembles resulting from this edge spreading were shown to have increased minimum distance growth rates and BEC thresholds when compared to equal rate ensembles from the family defined in Example 1 (see [8]). For example, for $L = 2$, $R_2 = 1/4$, $\delta_{min}^{(2)} = 0.0920$, and $\varepsilon^* = 0.6471$. The improved minimum distance growth rates are reflected in the upper bound on codes chosen from the QC sub-ensemble. For this family, we calculate $d_{minQC} \leq 176$ for $L \geq 2$.

*Example* 3: We now consider a 'bad' example of edge spreading. Consider the following component matrices obtained by edge spreading the all-ones base matrix $\mathbf{B}$ of size $3 \times 6$:

$$\mathbf{B}_0 = \begin{bmatrix} 1 & 1 & 1 & 0 & 0 & 0 \\ 1 & 1 & 1 & 0 & 0 & 0 \\ 0 & 0 & 0 & 1 & 1 & 1 \end{bmatrix} \text{ and } \mathbf{B}_1 = \mathbf{B} - \mathbf{B}_0.$$

This ensemble has relatively poor iterative decoding thresholds and minimum distance growth rates compared to the other asymptotically regular $(3,6)$ families. The BEC thresholds for this family converge to $0.4734$ (compared to $0.4881$ for the other asymptotically regular $(3,6)$ examples), and for $L = 2$, when $R_2 = 1/4$, the minimum distance growth rate is just $\delta_{min}^{(2)} = 0.0296$ with threshold $\varepsilon^* = 0.4949$. When calculating the upper bound on the minimum distance of members of the QC sub-ensemble $\xi^{QC}_{\mathbf{B}_{[0,L-1]}}(N)$ for this edge spreading, we note that, for any termination factor $L$, after some row permutations the ensemble contains the following sub-structure:

$$\boxed{\begin{matrix} 1 & 1 & 1 & 0 & 0 & 0 \\ 1 & 1 & 1 & 1 & 1 & 1 \\ 1 & 1 & 1 & 1 & 1 & 1 \\ 0 & 0 & 0 & 1 & 1 & 1 \end{matrix}},$$

which limits the circulant minimum distance to $d_{minQC} \leq 36$. This small upper bound for the QC sub-ensemble reflects the poor ensemble minimum distance growth rates.

*Example* 4: As a final asymptotically regular $(3,6)$ example, the edge spreading [8]

$$\mathbf{B} = \begin{bmatrix} 3 & 3 \end{bmatrix} \rightsquigarrow \mathbf{B}_0 = \begin{bmatrix} 2 & 1 \end{bmatrix} \text{ and } \mathbf{B}_1 = \begin{bmatrix} 1 & 2 \end{bmatrix}.$$

was shown to result in a family with the largest minimum distance growth rates of all the asymptotically regular $(3,6)$ families considered. In addition, for small values of $L$, the thresholds were shown to be the same as or larger than other asymptotically regular $(3,6)$ ensembles of the same rate, and, as with the other 'good' edge spreadings, the BEC BP thresholds converge to the optimal MAP decoding thresholds for $(3,6)$-regular ensembles ($\varepsilon = 0.4881$).

In this case, for $L = 2$, $R_2 = 1/4$, $\delta_{min}^{(2)} = 0.0950$, and $\varepsilon^* = 0.6447$. However, we note that for the circulant sub-ensemble, we obtain only $d_{minQC} \leq 30$ for $L \geq 2$, a relatively small upper bound, which can be achieved for small circulant size $N$. The parity-check matrix given as an example in Section III-A is a member of the QC sub-ensemble $\xi_{\mathbf{B}_{[0,1]}}(N)$. Using circulants of size $N = 38$ in this parity-check matrix, we achieve $d_{min} = 30$ (this is a $[152, 38, 30]$ binary linear code with girth 8).

We now show that by taking $m$-covers of this protograph we can increase the bound. For example, consider the following 2-cover:

$$\mathbf{B}'_0 = \begin{bmatrix} 1 & 1 & 1 & 0 \\ 1 & 1 & 0 & 1 \end{bmatrix} \text{ and } \mathbf{B}'_1 = \begin{bmatrix} 1 & 0 & 1 & 1 \\ 0 & 1 & 1 & 1 \end{bmatrix}.$$

Using these component submatrices, we obtain the base matrix $\mathbf{B}'_{[-\infty,\infty]}$ for a $(3,6)$-regular LDPC convolutional code ensemble with syndrome former memory $m_s = 1$. The terminated ensemble constructed from the component two-covers is denoted as $\xi_{\mathbf{B}'_{[0,L-1]}}(N)$. It follows that $\xi_{\mathbf{B}'_{[0,L-1]}}(N) \subseteq \xi_{\mathbf{B}_{[0,L-1]}}(2N)$, because any $N$-cover of an $m$-cover exists in the set of $mN$-covers of the original protograph. Interestingly, we calculate the minimum distance growth rate $\delta^{(2)}_{min} = 0.095$ (and threshold $\varepsilon^* = 0.6447$) for both the original ensemble $\xi_{\mathbf{B}_{[0,L-1]}}(N)$ and the two-cover ensemble $\xi_{\mathbf{B}'_{[0,L-1]}}(N)$. From this we conclude that typical codes with the same length from either ensemble would have the same minimum distance.

This is clearly not the case for the QC sub-ensembles. It is a simple exercise to choose circulants so that a code $C'$ from the quasi-cyclic sub-ensemble of the two-cover $\xi^{QC}_{\mathbf{B}'_{[0,L-1]}}(N)$ does not exist in the original QC sub-ensemble $\xi^{QC}_{\mathbf{B}_{[0,L-1]}}(2N)$, and vice versa. Using the 2-cover component submatrices, the upper bound on the minimum distance of members of the QC sub-ensemble increases to $d_{minQC} \leq 82$. The improvement can be verified quickly, since it is relatively easy to construct a code with minimum distance larger than 30 from this sub-ensemble.

Moreover, by taking 3-covers of the component submatrices:

$$\mathbf{B}''_0 = \begin{bmatrix} 1 & 1 & 0 & 1 & 0 & 0 \\ 1 & 0 & 1 & 0 & 1 & 0 \\ 0 & 1 & 1 & 0 & 0 & 1 \end{bmatrix}, \mathbf{B}''_1 = \begin{bmatrix} 1 & 0 & 0 & 1 & 0 & 1 \\ 0 & 1 & 0 & 1 & 1 & 0 \\ 0 & 0 & 1 & 0 & 1 & 1 \end{bmatrix},$$

the resulting terminated ensembles also have minimum distance growth rate $\delta^{(2)}_{min} = 0.095$ and threshold $\varepsilon^* = 0.6447$, yet we calculate $d_{minQC} \leq 210$ for codes from $\xi^{QC}_{\mathbf{B}''_{[0,L-1]}}(N)$ with $L \geq 2$. Comparing the value obtained for this 3-cover with Examples 2 and 3, which also have component submatrices of size $3 \times 6$ and $m_s = 1$, we obtain the largest bound for the ensemble with the largest minimum distance growth rate. The improvement we observe by taking graph covers of the protograph can be attributed to permitting arrays of circulants to replace entries in the base matrix $\mathbf{B}$.

Table I gives a summary of the results for Examples 1-4 considered above.

| Example | $\delta_{min}$ $(R = 1/4)$ | $\varepsilon^*$ $(R = 1/4)$ | Upp. bnd. on $d_{minQC}$ |
|---|---|---|---|
| 1 | 0.0815 | 0.6353 | 56 |
| 2 | 0.0920 | 0.6471 | 176 |
| 3 | 0.0296 | 0.4949 | 36 |
| 4 (3-cover) | 0.0950 | 0.6447 | 210 |

TABLE I: Comparison of $\delta_{min}$, BEC thresholds, and bounds on $d_{minQC}$ for several asymptotically regular $(3,6)$ families

## VI. Conclusions

Asymptotically regular LDPC codes based on protographs have been shown to display capacity approaching iterative decoding thresholds with minimum distance that grows linearly with block length. Both the minimum distance growth rate and threshold have been shown to depend closely on the choice of component protographs. In the interests of efficient implementation, this paper has explored the properties of the quasi-cyclic sub-ensembles of protograph-based codes. It was shown that, even though the members of the QC sub-ensemble are not typical members of the ensemble, the upper bound on the minimum Hamming distance of members of this sub-ensemble can be improved using choices of edge spreading that result in good ensemble minimum distance growth rates. In addition, the upper bound obtained for several of the examples presented here was shown to be achieveable by constructing codes with this minimum distance. Finally, we showed that the upper bound obtained for the QC codes in the ensemble can be improved by using arrays of circulants in a graph cover of the protograph. Due to space limitations, we have only presented results for edge spreadings of $(3,6)$-regular base matrices $\mathbf{B}$; however, similar results are observed for arbitrary $J$ and $K$.